\documentclass[twocolumn]{aastex631}
\usepackage{multirow}

\begin{document}

\title{Obscured at the Core: Evidence for Nuclear Dust in Reddened Type-1 AGN}

\author[0000-0002-0273-5890]{Miguel A. Montalvo Hernandez}
\affiliation{Department of Astrophysical Sciences, Princeton University, Princeton, NJ 08540, USA}

\author[0000-0003-4700-663X]{Andy D. Goulding}
\affiliation{Department of Astrophysical Sciences, Princeton University, Princeton, NJ 08540, USA}

\author[0000-0002-5612-3427]{Jenny E. Greene}
\affiliation{Department of Astrophysical Sciences, Princeton University, Princeton, NJ 08540, USA}

\begin{abstract}

Reddened Type-1 quasars offer a unique window into the structure and evolution of active galactic nuclei (AGN), yet their physical origin and the source of their reddening remain uncertain. Optical surveys often miss these dust-obscured objects, resulting in an incomplete view of the quasar population. In this work, we construct a sample of 6,600 Type-1 quasars at redshifts $\mathrm{0.5 \leq z \leq 1.2}$ by combining deep optical imaging from HSC with mid-infrared photometry from WISE, enabling a more complete selection that is not biased against reddened objects. We perform detailed SED modeling using the \texttt{CIGALE} code, enhanced by synthetic photometry derived from SDSS spectra to better constrain the optical continuum. We classify quasars into blue and reddened Type-1 populations based on their continuum slopes and compare their SEDs and emission line properties. As expected from this definition, reddened Type-1 AGN show higher dust extinction, with a median $\mathrm{A_V=0.60^{+0.32}_{-0.19}~mag}$, compared to $\mathrm{A_V=0.06^{+0.10}_{-0.03}~mag}$ for blue objects. But they also exhibit smaller torus half-opening angles, with a median of $25.7^{+10.1}_{-8.7}$~deg, compared to $33.3^{+11.1}_{-5.9}$~deg for blue objects. While such extinction could arise on either galaxy or nuclear scales, the systematically stronger narrow-line equivalent widths and weaker Balmer broad lines in reddened Type-1s indicate that the obscuration acts on nuclear scales, likely from dust concentrated near the polar axis. We discuss the possibility that these structural differences may be linked to a sub-pc outflow, that carries dusty gas into the polar region and evacuates the torus region.
\end{abstract}


\section{Introduction} \label{sec:intro}

Active galactic nuclei (AGN) are powered by the accretion of gas onto supermassive black holes (SMBH), releasing extraordinary amounts of energy across the electromagnetic spectrum \citep{salpeter1964,lynden1969}. In our standard AGN picture, this energy originates primarily from the SMBH’s accretion disk, which produces bright optical and ultraviolet (UV) continuum emission. Rapidly moving gas clouds near the SMBH, within the broad-line region, create broad optical and UV emission lines. Beyond the accretion disk and broad-line region lies an axisymmetric, clumpy torus of obscuring dust and gas, while slower-moving gas clouds outside this torus generate narrower emission lines. The standard unification model \citep{antonucci1993,urry1995,netzer2015} explains the observed differences between Type-1 and Type-2 AGN as a result of the viewing angle: a direct line of sight to the central source reveals the broad emission lines of Type-1 AGN, also known as broad-line AGN. An obscured line of sight through the dusty torus blocks the broad emission lines in Type-2 or narrow-line AGN, leaving only the narrower emission lines and infrared (IR) emissions from reprocessed radiation.

A subset of dust-obscured Type-1 AGN (also known as red quasars) exhibit redder optical \citep{richards2003} and near-IR colors than typical Type-1 AGN, primarily attributed to dust attenuation along the line of sight \citep{webster1995}. Although the unification model suggests that moderate inclination angles might lead to partial obscuration by the clumpy torus \citep{wilkes2002,rose2013,ananna2022A}, extreme cases of reddening may instead arise from obscuration on galactic scales, as suggested by \citealt{goulding2012}, where an edge-on view of the AGN host galaxy can block a significant fraction of the central light.

An alternative explanation for red quasars comes from merger-driven galaxy evolution. In this scenario, mergers between gas-rich galaxies trigger both star formation and AGN activity, with the central black hole growing rapidly while being heavily obscured by dust and gas within the host galaxy \citep{sanders1988,canalizo2001,hopkins2006}. Feedback from the SMBH eventually clears away this material, transitioning the quasar from obscured to unobscured. This model is supported by simulations of major mergers \citep{menci2004,hopkins2005,hopkins2006,hopkins2008} and observational evidence that red quasars exhibit higher fractions of merger features \citep{urrutia2008,glikman2015}, increased star formation activity \citep{georgakakis2009}, and higher luminosities (when corrected for extinction) at every redshift compared to unobscured quasars \citep{glikman2012}, as well as high accretion rates ($\mathrm{L/L_{Edd}\geq0.3}$, \citealt{kim2015,kim2024}). Despite these findings, the precise origin of the reddening, whether due to dust on nuclear scales, galaxy-wide obscuration, or a combination of both, remains unresolved. For a comprehensive review of obscuration mechanisms in AGN, see \cite{hickox2018}. Recent observations of the enigmatic population of Little Red Dots (LRDs; \citealt{labbe2025,barro2024,matthee2024}) at $\mathrm{z\geq4}$ helped motivate our investigation into reddening mechanisms in lower-redshift Type-1 quasars. LRDs are characterized by a steep red continuum in the rest-frame optical, along with prominent broad Balmer lines \citep{kocevski2023,matthee2024,greene2024}, suggesting AGN activity.

Building a large sample of blue to reddened broad-line AGN offers a unique opportunity to investigate the physical origin of obscuration in Type-1 AGN. By modeling their SEDs from near-UV to mid-IR, we can constrain key properties of the dusty torus and evaluate its role in producing the observed reddening. Established SED modeling frameworks (e.g., \citealt{pier1992}; \citealt{fritz2006}; \citealt{nenkova2008}; \citealt{Stalevski2012,Stalevski2016}; \citealt{tanimoto2019}) allow us to accurately account for AGN emission, while simultaneously incorporating contributions from stellar populations and dust within the host galaxy (e.g., \citealt{calistro2016}; \citealt{leja2018}; \citealt{boquien2019}). This comprehensive approach enables us to disentangle the relative contributions of nuclear- and galactic-scale dust to the overall obscuration. For instance, if reddening correlates with parameters such as the torus half-opening angle or inclination, this would support a torus-scale origin. Conversely, if the SEDs indicate additional contributions from polar dust or large-scale host galaxy obscuration, this would align with scenarios involving galaxy mergers.

In this paper, we present a detailed study of the spectral energy distributions (SEDs) of Type-1 quasars, focusing on the physical origin of their reddening. In Section \S \ref{sec:sample} we describe the sample selection and the photometric and spectroscopic data used. Section \S \ref{sec:methods} outlines our methodology used for synthetic photometry and SED modeling. Section \S \ref{sec:results} presents our SED fitting results, including the classification of blue and reddened Type-1 AGN and the analysis of their composite spectra. Section \S \ref{sec:discussion} discusses the implications of our findings and Section \S \ref{sec:conclusions} provides the final conclusions of this work. Throughout this study, we adopt the standard $\Lambda$CDM cosmology with $\mathrm{H_0 = 70~km~s^{-1}~Mpc^{-1}}$, $\mathrm{\Omega_m = 0.3}$, and $\mathrm{\Omega_\Lambda = 0.7}$.

\section{Sample} \label{sec:sample}

Traditional quasar selection methods, such as those employed in the Sloan Digital Sky Survey (SDSS; \citealt{york2000}), rely on optical color cuts designed to identify luminous, unobscured quasars \citep{richards2002}. While effective at isolating broad-line AGN, these techniques introduce selection biases that systematically exclude dust-reddened or obscured quasars, leading to an incomplete census of AGN activity. Infrared (IR) photometry provides a more inclusive view of AGN \citep{glikman2004,lacy2004,stern2005,maddox2008,kirkpatrick2013}, as these sources emit strongly in the IR regardless of their dust properties. While optical surveys may miss dust-reddened quasars due to their faintness at shorter wavelengths, the combination of deep optical and mid-IR photometry can recover a more representative AGN population.

\begin{figure}[t]
    \centering
    \includegraphics[scale=0.52]{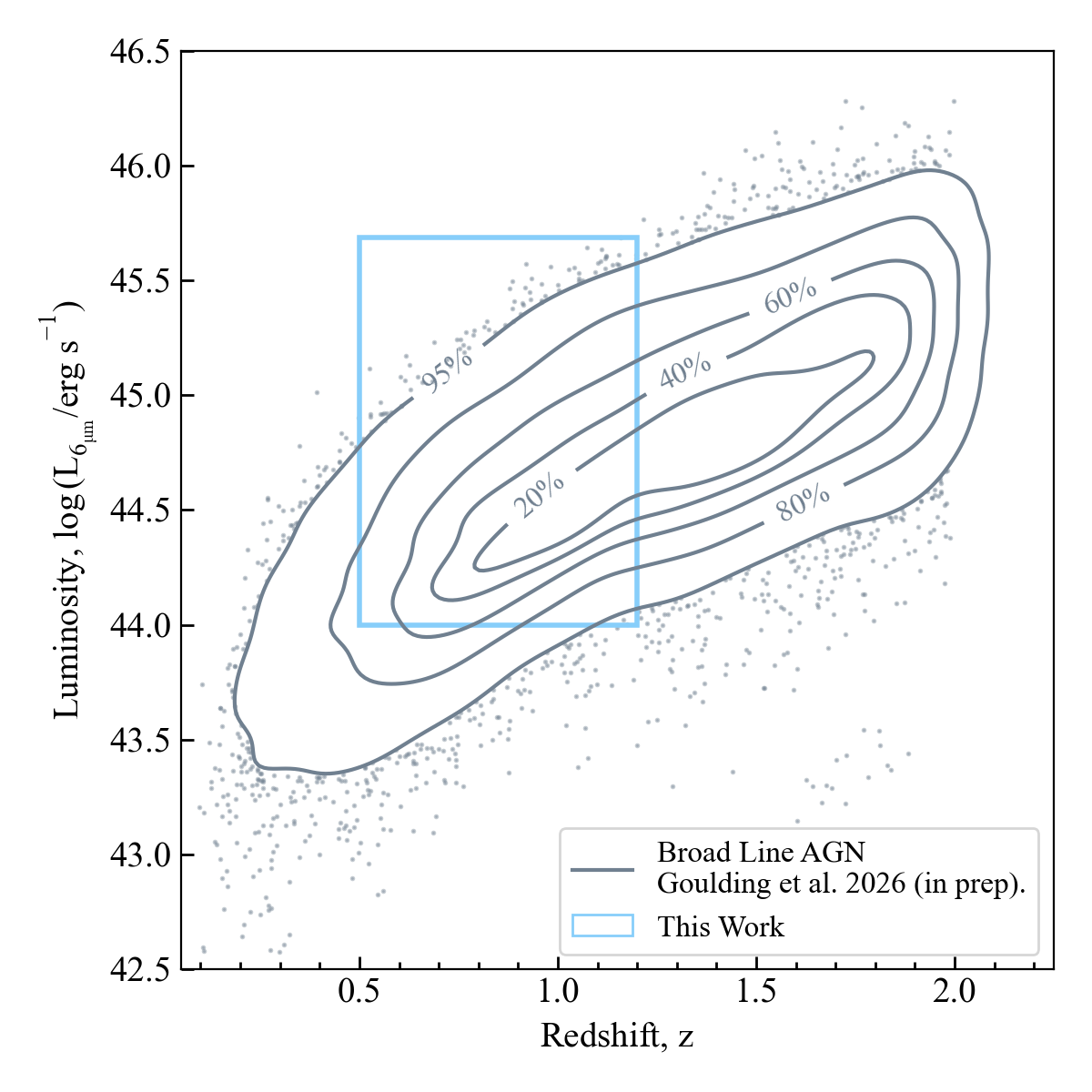}
    \caption{All broad line AGN sources in Goulding et al. 2026 (in prep.) cross-matched with SDSS spectra are shown in the gray contours. The subset used in this work, selected for completeness over $\mathrm{0.5 \leq z \leq 1.2}$ and $\mathrm{L_{6\mu m} \equiv \nu L_{\nu} \geq 10^{44}~erg~s^{-1}}$, is indicated by the light-blue box. Contours enclose 20, 40, 60, 80, and 95\% of the sample; individual objects outside the 95\% contour are shown as scatter points.
    \label{fig:sample}}
\end{figure}

To achieve this selection, we leverage a dataset that integrates optical photometry from the Hyper Suprime-Cam Subaru Strategic Program (HSC-SSP; \citealt{miyazaki2018}) with mid-IR measurements from the Wide-field Infrared Survey Explorer (WISE; \citealt{wright2010}). The HSC-SSP provides imaging in five broadband filters (g, r, i, z, and y) centered at wavelengths of 0.47, 0.62, 0.77, 0.89 and 0.98 $\mu$m, respectively, with a median $5\sigma$ depth of 25.1 mag \citep{aihara2018}. Meanwhile, WISE scanned the entire sky in four mid-IR broadband filters (W1, W2, W3, and W4), centered at 3.4, 4.6, 12, and 22 $\mu$m, respectively. The parent quasar sample used in this work (Goulding et al. 2026, in prep.) combines photometric data from these two key surveys. The depth of HSC-SSP allows us to detect even heavily obscured AGN that might be missed by shallower optical surveys, while the wide-area coverage of WISE ensures a statistically significant sample of quasars spanning a broad range of obscuration properties. 

Goulding et al. 2026 (in prep.) employ the powerful dimensionality reduction algorithm Uniform Manifold Approximation and Projection (UMAP; \citealt{mcinnes2018}) to disaggregate stars and non-active galaxies from AGN/quasar systems. They identify a unique clustered region of space in the projected 2-D manifold that predominantly corresponds to objects in the SDSS Quasar Catalog \citep{lyke2020}. In this work, we cross-match objects in this UMAP region with spectroscopy from the SDSS Sixteenth Data Release (DR16; \citealt{blanton2017,ahumada2020}) and further restrict the sample to objects within the redshift range $\mathrm{0.5 \leq z \leq 1.2}$ and with rest-frame $\mathrm{6\mu m}$ luminosities $\mathrm{L_{6\mu m} \equiv \nu L_{\nu} \geq 10^{44}~erg~s^{-1}}$, to ensure sample completeness (see Figure \ref{fig:sample}). This luminosity criterion ensures that our study concentrates on luminous quasars, yielding a robust dataset of 6,600 quasar-like objects for subsequent analysis.

Of these, Sloan spectra identify broad-line features in approximately three-quarters of them. The remaining quarter remains ambiguous from SDSS fits. To assess this subset, we visually inspect about one-third of the ambiguous cases and find that 99\% exhibit broad-line features, while the remaining 1\% have spectra with signal-to-noise ratios too low to confidently determine their classification. This suggests that nearly all sources in our sample are broad-line AGN, with only about 1\% that either lack broad-lines or are too faint to determine their spectral class. \citealt{hviding2024} followed up a similar subsample from the same parent sample (Goulding et al. 2026, in prep.) with deeper spectroscopy and confirmed their broad-line nature.

This innovative approach leverages the deep imaging capabilities of HSC and the extensive coverage of WISE, enhanced by unsupervised machine learning techniques, to achieve a higher number density AGN sample than in previous studies. The final sample consists of 6,600 sources with a median i-band magnitude of $20.2$.

\begin{figure*}
    \centering
    \includegraphics[width=\textwidth]{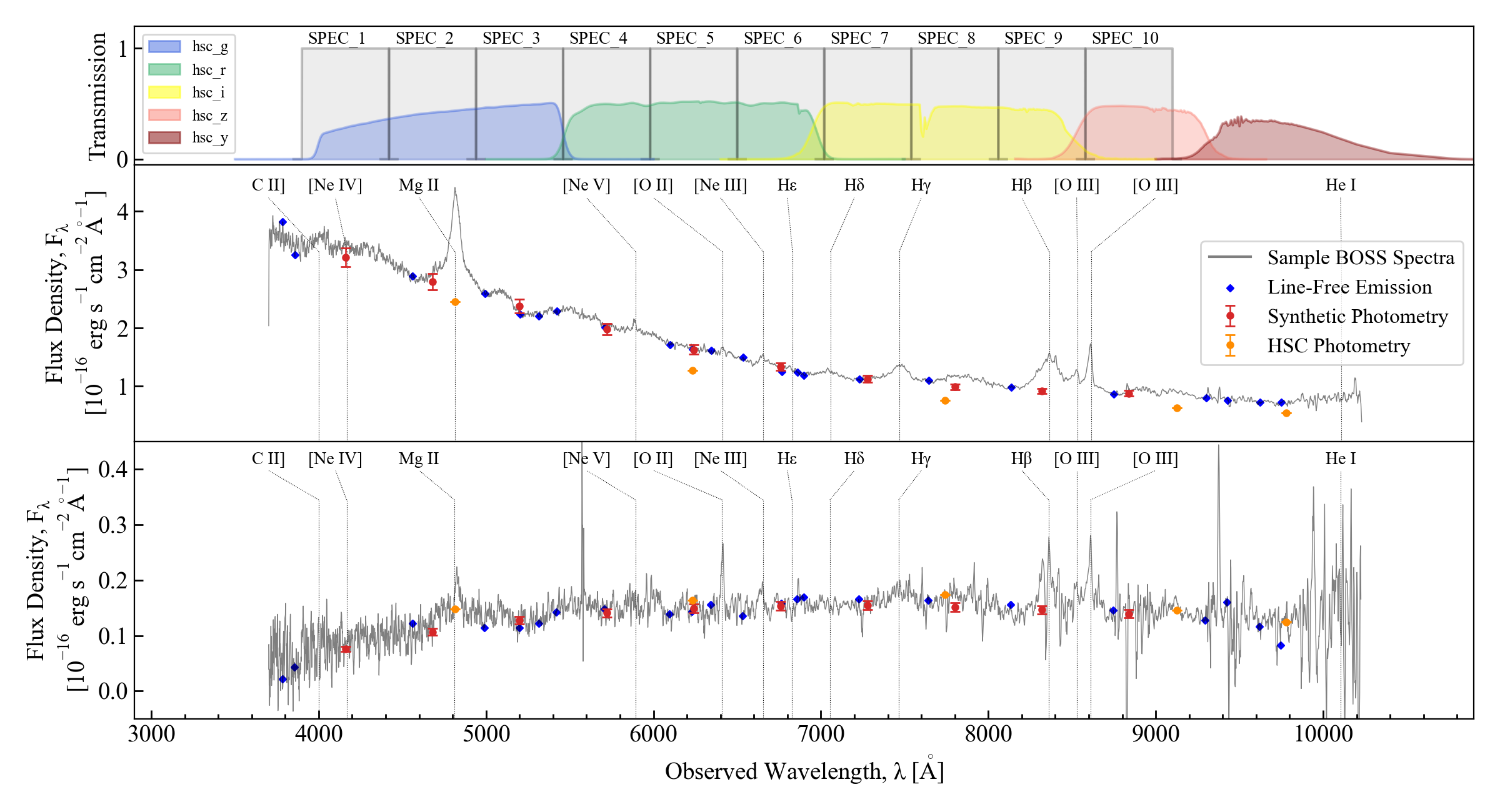}
    \caption{\textbf{Top:} HSC filter transmission curves (g,r,i,z,y) along with the wavelength coverage of ten flat synthetic filters that we use to perform synthetic photometry on the SDSS spectra. \textbf{Middle:} Sample spectrum of a typical Type-1 AGN in our sample: SDSS J022841.08+003049.4. \textbf{Bottom:} Sample spectrum of a reddened Type-1 AGN in our sample: SDSS J222936.55+032039.3. In the middle and bottom panels we plot the smoothed SDSS spectra (in gray) along with their HSC photometry (in orange) and the calculated synthetic photometry obtained from their spectra (in red). The selected wavelengths, free of emission lines, that trace the continuum emission are plotted in blue. Prominent narrow and broad emission lines are shown as well (dotted).
    \label{fig:spectra}}
\end{figure*}

\section{Methods} \label{sec:methods}

To understand the physical properties of the quasars, we aim to fit their SEDs from the UV to the mid-IR. To remove the impact of emission lines from the photometry for this fitting, we begin by performing synthetic photometry using data derived from SDSS spectra. This cleaned photometry is then used as input for the SED fitting process.

\subsection{Synthetic Photometry} \label{subsec:photometry}

To accurately model the spectral energy distributions of our sample, spanning from rest-frame UV-optical to mid-IR, we aim to capture the intrinsic AGN continuum while disentangling it from host galaxy emission and accounting for reddening effects. Given that HSC photometry is limited to five broadband filters, we enhance the resolution of our optical SEDs by incorporating synthetic photometry derived from SDSS spectra. This method provides a denser sampling of the continuum, allowing us to better constrain its shape and minimize deviations caused by emission lines, thereby improving sensitivity to reddening at shorter wavelengths (See Figure \ref{fig:spectra}). By leveraging the spectral data, we can discern subtle variations in the SEDs that are crucial for distinguishing between AGN and host galaxy contributions.

To implement this approach, we define ten synthetic box filters ranging from $3900~\mathrm{\mathring{A}}$ to $9100~\mathrm{\mathring{A}}$ in the rest frame, which corresponds to the wavelength range common to all objects in our subsample with observations from the BOSS and SDSS surveys. To accurately trace the shape of the AGN continuum emission, we select approximately 30 wavelengths free of emission lines, based on the composite spectrum from \citet{vandenberk2001}. The flux at each line-emission-free wavelength is obtained by averaging the spectrum within a $30~\mathrm{\mathring{A}}$ interval centered on the selected position. Note that these wavelengths still include contributions from the broad Fe II and Fe III emission complexes.

To determine the uncertainties in the flux densities, we perform 500 random draws within $\pm10~\mathrm{\mathring{A}}$ of the emission-free wavelengths that fall within the observed spectrum. For each realization, we fit a spline to these points to estimate the continuum shape, and calculate the flux density for each synthetic filter by averaging the spline fit within the corresponding filter range. The final synthetic photometry is obtained by averaging the flux densities from the 500 realizations. The first component of the uncertainty is given by the standard deviation of the flux densities from these 500 spline-fitted models. To account for systematic effects and prevent unrealistically small errors, we adopt a second component corresponding to a 5\% flux floor. The final uncertainties are obtained by adding in quadrature the standard deviation from the 500 realizations and this 5\% flux floor.

This synthetic photometry approach aids in the separation of AGN and host galaxy contributions within the optical region of the SED. By increasing the optical continuum resolution, we can distinguish features indicative of early-type versus star-forming galaxies. Figure \ref{fig:spectra} shows the results of synthetic photometry performed on the SDSS spectrum of a typical Type-1 AGN, SDSS J022841.08+003049.4, and a reddened Type-1 AGN, SDSS J222936.55+032039.3, in our sample. As seen, this technique effectively traces the continuum shape while avoiding deviations caused by emission lines and better capturing the effects of reddening at short wavelengths.

\subsection{SED Modeling} \label{subsec:sed}

We now have ten synthetic photometry points in the UV-optical range that are free of line-emission contributions and four photometry points in the mid-IR range, which we use to fit the SED of each source. To fit the SED, we employ the Python-based Code for Investigating GALaxy Emission (\texttt{CIGALE}; \citealt{burgarella2005,noll2009,boquien2019}), in its latest version, \texttt{CIGALE V2022.0} (\citealt{yang2020,yang2022}). The code allows input of all relevant parameters to generate a grid of model SEDs, and then computes the likelihood-weighted mean of these quantities by Bayesian estimation. \texttt{CIGALE} conserves the energy balance between emission absorbed by dust and its thermal re-emission. To model the primary components of AGN emission, including the accretion disk and the dusty torus, we use the UV-to-IR SED model of AGN, \textsc{SKIRTOR}. \citealt{Stalevski2012,Stalevski2016} describe the model in detail but we provide a summary below.

\begin{deluxetable*}{ccc}
\tablecaption{Parameters used in \texttt{CIGALE} to model the UV-optical to mid-IR SED of Type-1 AGN. \label{tab:params}}
\tablehead{ \colhead{Parameter} & \colhead{Description} & \colhead{Values} }
\startdata
\midrule
\multicolumn{3}{c}{Delayed SFH} \\ \midrule
$\mathrm{\tau_{main}}$ & e-folding time of the main stellar population model in Myr & 1000, 2000, 4000 \\
$\mathrm{t_{main}}$ & Age of the main stellar population in the galaxy in Myr & 4000, 6000, 8000 \\
$\mathrm{f_{burst}}$ & Mass fraction of late burst population & 0.0 \\ \midrule
\multicolumn{3}{c}{Simple Stellar Population \citep{bruzual2003}} \\ \midrule
IMF & Initial mass function & Salpeter (\citeyear{salpeter1955}) \\
Z & Metallicity  & 0.004, 0.02 \\ 
$\mathrm{t_{separation}}$ & Age of separation between young and old stellar populations & 4000, 6000, 8000 \\ \midrule
\multicolumn{3}{c}{Galactic Dust Attenuation \citep{charlot2000}} \\ \midrule
$\mathrm{A_{V}^{ISM}}$ & V-band attenuation in the interstellar medium & 0.1, 0.35, 0.65, 1.0 \\
$\mathrm{\delta_{ISM}}$ \& $\mathrm{\delta_{BC}}$ & Power law slope of the attenuation curves & -0.7 \& -1.3 \\ \midrule
\multicolumn{3}{c}{Galactic Dust Emission \citep{draine2007,draine2014}} \\ \midrule
$\alpha$ & Slope in $\mathrm{dM_{dust}\propto U^{-\alpha}dU}$ & 2.0 \\
$\gamma$ & Mass fraction of dust linked to the star-forming regions & 0.1 \\ \midrule
\multicolumn{3}{c}{AGN Emission \citep{Stalevski2016}} \\ \midrule
$\Theta$ & Half-opening angle measured between equatorial plane and edge of torus & $10^\circ$, $20^\circ$, $30^\circ$, $40^\circ$, $50^\circ$, $60^\circ$, $70^\circ$ \\
$i$ & Inclination angle of the polar axis relative to the line of sight & $10^\circ$, $20^\circ$, $30^\circ$, $40^\circ$ \\
\multirow{2}{*}{$\mathrm{E(B-V)}$} & \multirow{2}{*}{Dust extinction in the polar direction} & 0.005, 0.01, 0.05, 0.1, 0.15, 0.2, 0.25, \\
& & 0.3, 0.35, 0.4, 0.5, 0.6, 0.9 mag \\
$\mathrm{\tau_{9.7}}$ & Average equatorial optical depth at $9.7\mu m$ & 5, 7, 9 \\
\enddata
\tablecomments{For parameters not listed here, we use the default values.}
\end{deluxetable*}

Continuum emission from the disk is modeled by a piecewise set of power-laws as in \citealt{Stalevski2016} with four components that cover $0.008\mu m$ to $1000\mu m$. \texttt{CIGALE V2022.0} introduces a flexible power-law index that modifies its original value in the range $0.1\mu m$ to $5\mu m$ to account for intrinsic dispersion in UV/Optical slopes. We use the default value of $-0.36$. The emission from the accretion disc is anisotropic, being strongest perpendicular to the disc and absent in the equatorial plane.

The dusty torus is modeled as a flared disc characterized by an inner radius ($R_{in}$) and an outer radius ($R_{out}$). In the context of the \textsc{SKIRTOR} model, the torus height is parameterized by the "half-opening angle" ($\Theta$), which is defined as the angle measured from the equatorial plane to the edge of the torus. A large half-opening angle means a thicker torus with a large torus height. A small half-opening angle means a thin torus with a small torus height. We will adopt this convention for consistency. The inner radius is set by the dust sublimation temperature, which depends on the central source luminosity, while the outer radius marks the extent of the torus. We assume a fixed ratio of outer to inner radius, $R_{out}/R_{in}=20$, and allow the half-opening angle to vary freely in our model, as it is one of the key parameters influencing the SED that our photometric data can constrain.

Dust in the torus is modeled as a two-phase medium, consisting of numerous high-density clumps dispersed within a smooth, low-density component. The clumpiness of the torus is defined by the total number of clumps and the fraction of dust mass within these clumps ($f_{cl}$). The model sets this fraction as 97\% of the total dust mass being inside clumps and only 3\% in the diffuse dust between clumps. The average equatorial optical depth at $9.7\mu m$ ($\tau_{9.7}$) determines the overall opacity, influencing the radiation absorbed and scattered by dust. Due to the clumpy medium, the optical depth varies along different sight-lines and given that we can't constrain this parameter very well, as shown in \citealt{yang2020}, we only allow for three possible values centered around 7, to be in accordance with their work. The dust density follows a power law in both the radial direction and polar angle, with the steepness of the gradient determined by power law indices $p$ and $q$ for the radial and polar directions, respectively. We assume $p=q=1$ in our model. 

\begin{figure}[t]
    \centering
    \includegraphics[scale=0.49]{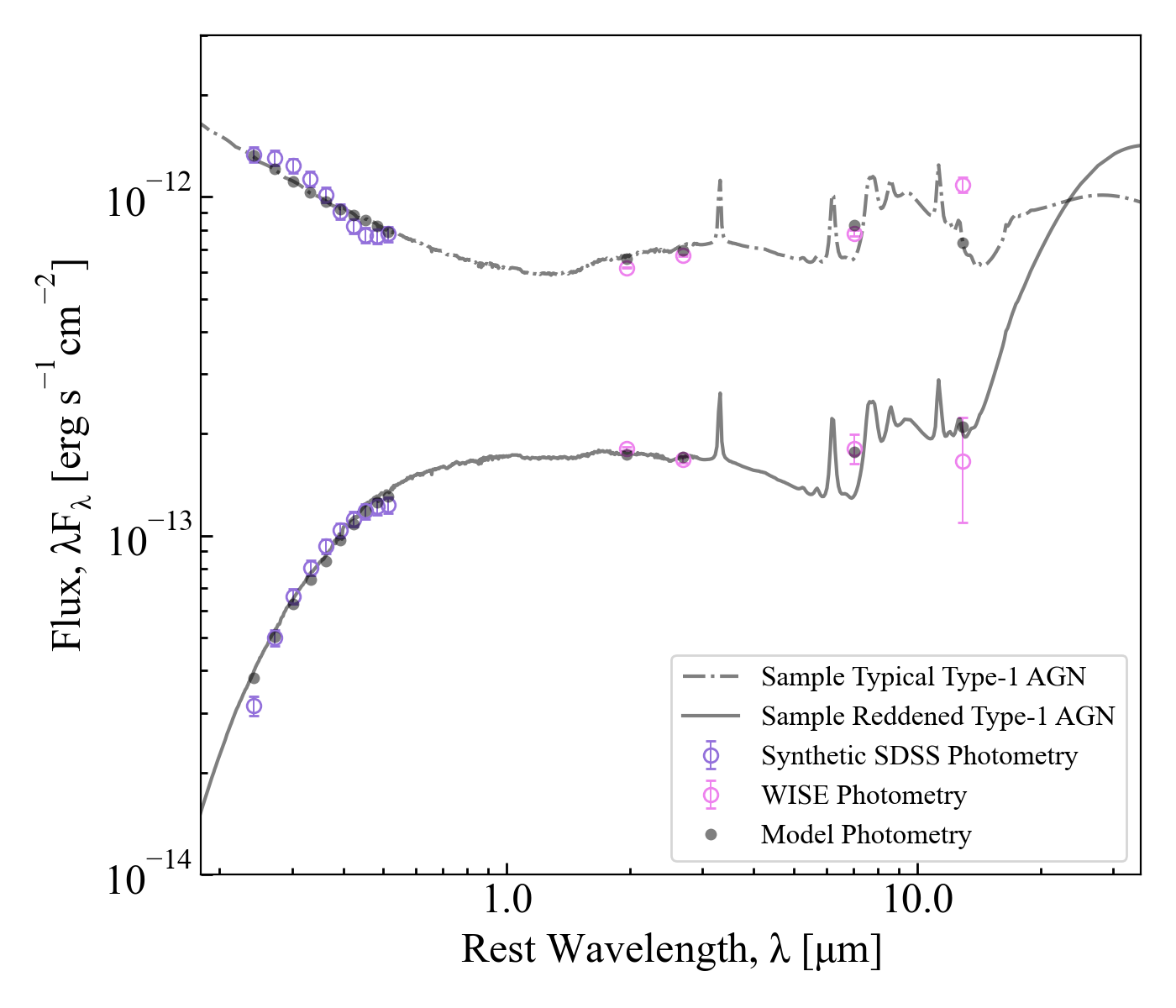}
    \caption{Sample SED fits of a typical (dot-dash line) and a reddened (solid line) Type-1 AGN in our sample. We plot the synthetic optical photometry obtained from the SDSS spectrum (purple circles) and the WISE photometry (pink circles) on top of the fits, along with the best fit model photometry (black dots). These two sources are at the same redshift ($\mathrm{z=0.72}$), highlighting the differences in their SED shapes. The typical Type-1 object, SDSS J022841.08+003049.4, has a fit with reduced $\mathrm{\chi^2=1.02}$ and the reddened Type-1, SDSS J222936.55+032039.3, has a fit with reduced $\mathrm{\chi^2=0.46}$.}
    \label{fig:sed}
\end{figure}

Another important parameter is the inclination angle ($i$) of the polar axis relative to the line of sight. When the line of sight is clear of the dusty torus, the radiation from the accretion disc is directly visible. However, when the line of sight intersects the dusty torus, most of the radiation is absorbed and re-emitted at longer wavelengths.  Since we have confirmed that our sources have broad lines, we choose a narrow range of values that correspond to Type-1 AGN, namely [$10^\circ$, $20^\circ$, $30^\circ$, $40^\circ$]. The \textsc{SKIRTOR} model does not include an additional line-of-sight dust component that attenuates the direct emission from the AGN accretion disk. Such a component could arise from dust in the polar direction, as suggested by interferometric observations, which reveal extended dust structures along the polar axis in some AGN \citep{asmus2016,stalevski2017}. To address this lack, \citet{yang2020} incorporates a "polar dust" component, which acts as a dust screen in front of the accretion disc, without specifying its radial location, and is characterized by various empirical extinction curves. We employ the Small Magellanic Cloud (SMC) extinction curve \citep{prevot1984} for our analysis, as it generally provides better fits for AGN SEDs \citep{hopkins2004, bongiorno2012}. The amplitude of polar dust extinction is a free parameter in our modeling. \citet{yang2022} maintains conservation of energy in this new component with the re-emission of polar dust modeled by a modified blackbody. The polar dust temperature is set to 100K \citep{buat2021} and its emissivity index is set to 1.6 \citep{casey2012} for our modeling.

For the galaxy component, we adopt a delayed$- \tau$ star formation history (SFH) model, with $\mathrm{SFR(t)\propto t~exp(-t/\tau)}$, where we parameterize the age ($\mathrm{t_{main}}$) and $e$-folding time ($\mathrm{\tau_{main}}$) of the main stellar population in the galaxy. We use the simple stellar population (SSP) models of \citealt{bruzual2003} with a Salpeter initial mass function (IMF; \citealt{salpeter1955}), while parameterizing the metallicity of the stellar model and the age of separation between young and old stellar populations. For dust attenuation to the stellar components, we used the model provided by \citet{charlot2000} that has two different power-law attenuation curves of the form $\mathrm{A(\lambda)\propto \lambda^{\delta}}$; one for the interstellar medium (ISM) and one for the birth clouds (BC). We set their slopes to the default values of $\mathrm{\delta_{ISM}=-0.7}$ and $\mathrm{\delta_{BC}=-1.3}$, and parameterize the V-band attenuation in the ISM ($\mathrm{A_V^{ISM}}$). The ratio of the total attenuation undergone by stars older than 10 Myr to that undergone by stars younger than 10 Myr, defined as $\mathrm{\mu=A_V^{ISM}/{({A_V^{BC}+A_V^{ISM}})}}$, is also set to the default value of $0.44$. 

The reprocessed IR dust emission of UV/optical  stellar radiation is modeled as in \citealt{draine2007}, including the improvements in \citealt{draine2014}. The model includes two components: The first is the diffuse dust emission, heated by the global stellar population, with mass fraction $\mathrm{(1-\gamma)}$, while the second is dust emission specifically linked to star-forming regions, with mass fraction $\mathrm{\gamma}$, where dust is heated by a variable radiation field ranging from $\mathrm{U_{min}}$ (set to 1.0) to $\mathrm{U_{max}=10^{7}}$. 
We tabulate all the free parameters for our model in Table \ref{tab:params}. An example of the SED fit output from \texttt{CIGALE V2022.0} is shown in Figure \ref{fig:sed}.

\begin{figure*}
    \includegraphics[width=\textwidth]{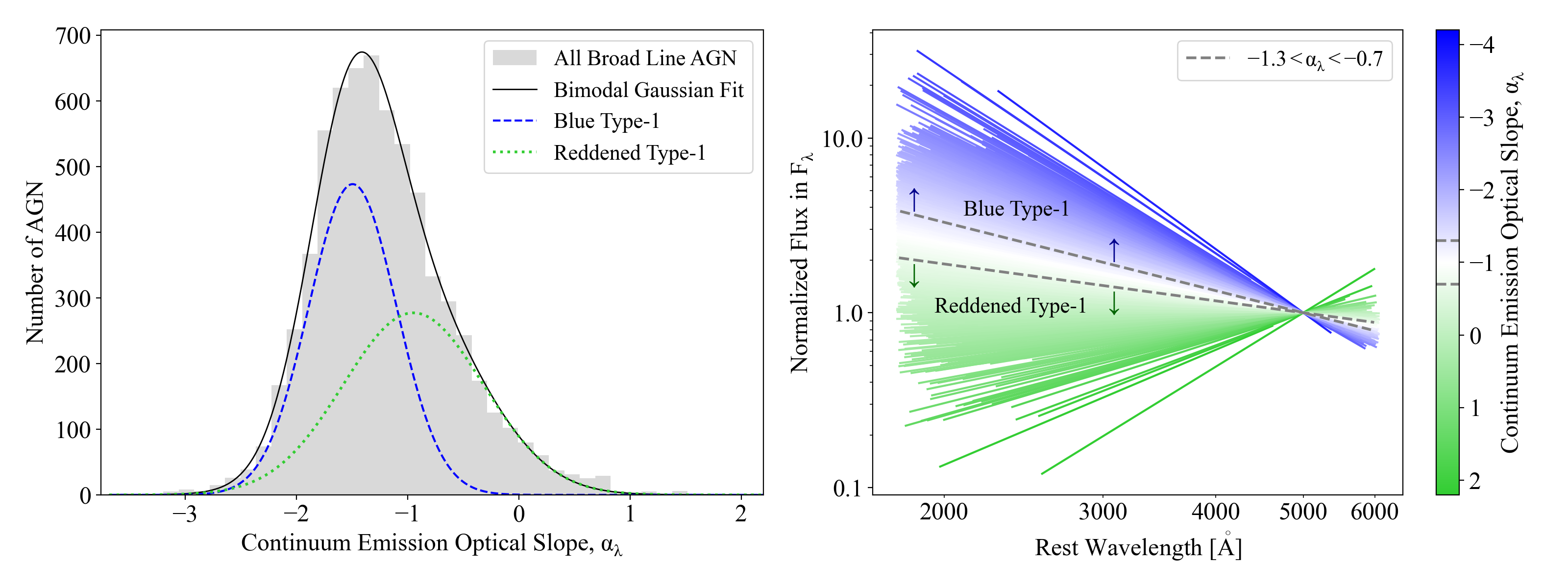}
    \caption{\textbf{Left:} Bimodal Gaussian fit (black solid) to the distribution of optical slope values, $\mathrm{\alpha_{\lambda}}$, of the continuum emission (gray histogram) in log-space for Type-1 AGN in our sample. The blue (dashed blue) and reddened (dotted green) Type-1 AGN distributions that form the bimodal Gaussian fit intersect at $\mathrm{\alpha_{\lambda}\sim-1}$. \textbf{Right:} Distribution of linear fits to the continuum emission in log-log space that shows our classification between blue and reddened Type-1 AGN (gradient from blue to green). The objects that lie within $\mathrm{-1.3<\alpha_{\lambda}<-0.7}$ are excluded from the classification (dotted black).
    \label{fig:alphas}}
\end{figure*}

\section{Results} \label{sec:results}

In this section, we present a comprehensive analysis of the spectral energy distributions (SEDs) of Type-1 quasars to investigate the physical origin of their reddening. We begin by classifying the broad-line objects into blue and reddened Type-1 AGN based on their optical continuum power-law slopes. Next, we construct composite spectra for each sub-sample to compare their large-scale continuum shapes. We then calculate the equivalent widths of prominent spectral lines to understand the differences in emission and absorption features between the two classes, which will also help showcase the differences in galaxy contributions. Following this, we describe the outputs from the SED fitting using \texttt{CIGALE}, focusing on key parameters such as the inclination angle, half-opening angle of the dusty torus, and dust reddening. Finally, we explore the relationship between the E(B-V) color excess and the half-opening angle of the torus, as illustrated in Figure 8, to gain insights into the structural differences between blue and reddened Type-1 AGN. 

\begin{figure*}
    \centering
    \includegraphics[width=\textwidth]{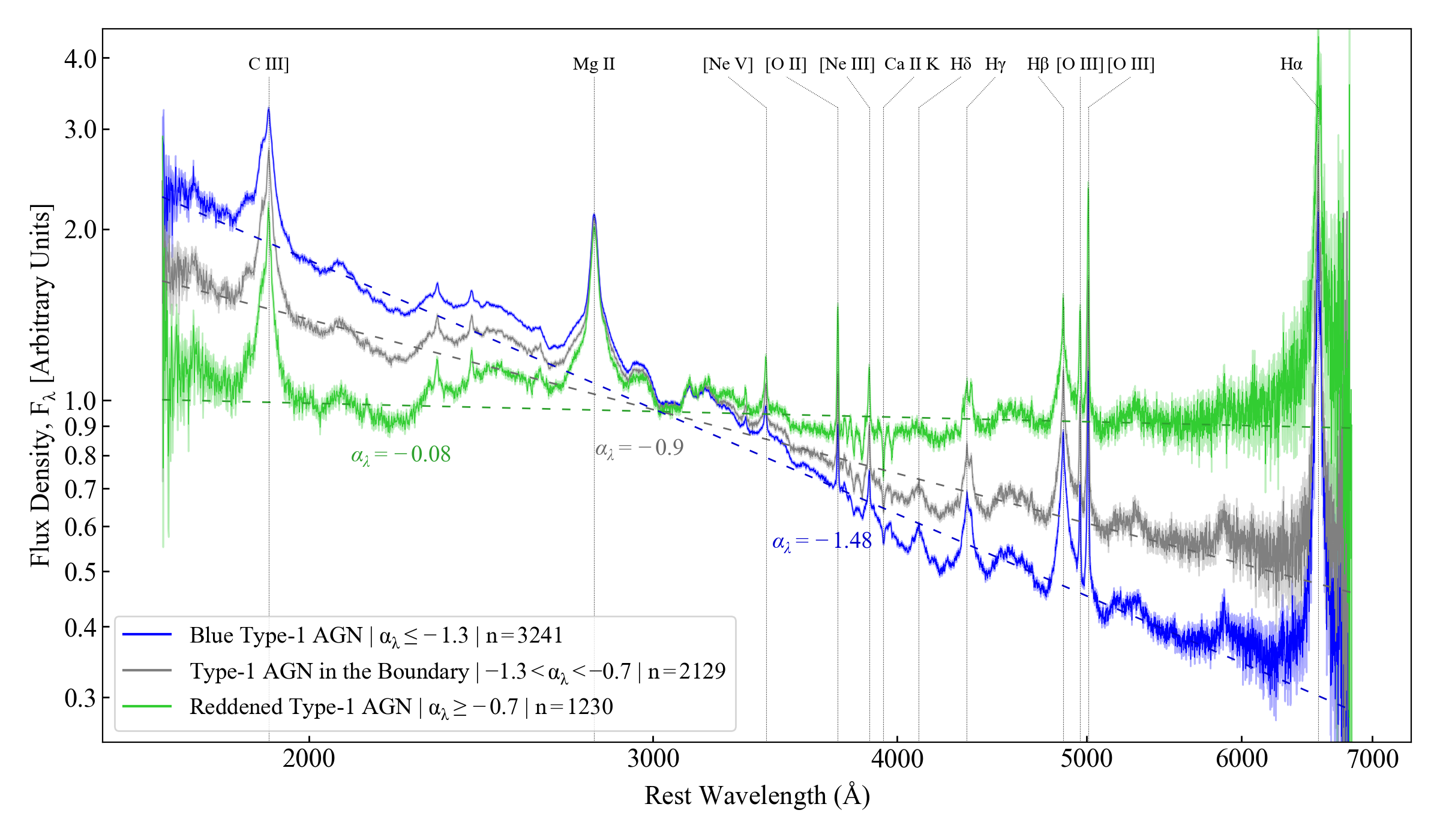}
    \caption{Composite spectra of blue Type-1 AGN (solid blue) and reddened Type-1 AGN (solid green) using a geometric mean to maintain the shape of the continuum emission along with their 68\% confidence level (light-blue and light-green). The linear fit to the blue Type-1 AGN continuum (dashed blue) has a slope of \(\alpha_{\lambda}=-1.48\) while the linear fit to the reddened Type-1 AGN continuum has a slope of \(\alpha_{\lambda}=-0.08\). Some prominent emission and absorption lines are displayed (grey dotted). The composite spectrum for the sources that lie in the boundary of definition between blue and reddened Type-1 AGN is also plotted (solid gray) with it's corresponding 1$\sigma$ uncertainty (light-gray).
    \label{fig:composite}}
\end{figure*}

\subsection{Classification of Blue and Reddened Type-1 AGN} \label{subsec:alphas}

The continuum emission from AGN can be approximated by a power law, $f_{\lambda} \propto \mathrm{\lambda^{\alpha_{\lambda}}}$, where $\alpha_{\lambda}$ is the exponent of the fit and corresponds to the slope of the continuum in log–log space. We characterize the distribution of these optical slopes, derived from power-law fits to the continuum normalized at $5000~\mathrm{\mathring{A}}$ rest-frame. The distribution does not exhibit a clear bimodality but instead forms a smooth continuum resembling a Gaussian with a tail toward redder colors. To quantify this red tail and define a practical division between the blue Type-1 and reddened Type-1 sources, we fit the histogram of $\alpha_{\lambda}$ with a sum of two Gaussian components. The approximate intersection of these two components, at $\alpha_{\lambda}\sim -1$, is used to separate blue Type-1 and reddened Type-1 AGN, as shown in Figure~\ref{fig:alphas}. Since the classification near this boundary is somewhat arbitrary, we exclude objects with slopes in the range $\mathrm{-1.3<\alpha_{\lambda}<-0.7}$ to ensure that our results for the two classes are robust. The resulting classification yields 3,241 blue Type-1 AGN and 1,230 reddened Type-1 AGN, while 2,129 sources lying within the boundary region are excluded from the analysis.

\subsection{AGN Composites} \label{subsec:composite}

We are interested in comparing the large-scale continuum shape of blue and reddened Type-1 AGN. To achieve this, we create a composite spectrum for each class using their SDSS spectra. We use the geometric mean to generate the composite and preserve the total continuum shape, following the same method as 
\citet{vandenberk2001}. The geometric mean is defined as \(\left\langle f_{\lambda} \right\rangle = \left( \prod_{i=1}^{n} f_{\lambda,i}\right)^{\frac{1}{n}}\), where \(f_{\lambda,i}\) is the flux density of the \(i\)-th spectrum in the bin centered on wavelength \(\lambda\), and \(n\) is the number of spectra contributing to the bin. 

Spectroscopic redshifts for each source come from the SDSS-DR16 quasar catalog \citep{lyke2020} where available. Otherwise, they are taken from the corresponding spectrum headers in SDSS-DR16 \citep{ahumada2020}. The spectra from each sub-sample are shifted to the rest frame using their spectroscopic redshift values, and then re-binned onto a common wavelength grid. We construct the grid based on the minimum and maximum values of the rest-frame wavelengths, with a resolution of 1 \(\mathrm{\mathring{A}}\) per bin. We resample using \texttt{SpectRes}: Simple Spectral Resampling tool \citep{carnall2017}, which cross-matches the original wavelength grid with the new one. If an original wavelength bin overlaps with multiple bins in the new grid, its flux density is redistributed proportionally based on the fractional overlap, ensuring that the total flux density is conserved. Subsequently, we normalize the rebinned spectra to unit average flux density over the rest-wavelength interval \(3020-3100~\mathrm{\mathring{A}}\), which is free of strong emission lines. The geometric mean of the flux density values is computed in each wavelength bin, producing the geometric mean composite AGN spectrum. To get an estimate of the uncertainty in the flux density, we perform a bootstrap analysis with 1000 realizations, sampling with replacement, and calculating the 68\% confidence level interval at each wavelength. The resulting composite spectra are displayed in Figure \ref{fig:composite} on a logarithmic scale.

Figure \ref{fig:composite} presents the composite optical spectra for blue Type-1 AGN (\(\alpha_{\lambda} \leq -1.3\)), reddened Type-1 AGN (\(\alpha_{\lambda} \geq -0.7\)), and objects that fall within the boundary between these definitions (\(-1.3 < \alpha_{\lambda} < -0.7\)), which are excluded from the analysis. The spectra are plotted as a function of rest wavelength, with prominent emission and absorption features labeled for reference. Each composite exhibits a power-law continuum indicative of AGN-dominated emission, accompanied by notable emission lines such as C III], Mg II, [O III], the Balmer series, and prominent Fe II emission.

We use the same 30 wavelengths that are emission-line free in the composite spectra, and then fit a straight line in log-log space to the flux densities at those locations, to get a value for the power-law index of the continuum emission in the optical. The blue Type-1 AGN spectrum has the steepest slope, characterized by a power-law index of \(\alpha_{\lambda} = -1.48\), and exhibits minor stellar absorption, suggesting minimal contribution from the host galaxy. In contrast, the reddened Type-1 AGN spectrum shows a much flatter slope (\(\alpha_{\lambda} = -0.08\)) and displays pronounced stellar absorption features, including Ca II H and K lines and higher-order Balmer lines, indicative of significant host galaxy contamination and/or dust reddening.

The differences between these composites raise the key question of what causes the redder continuum slope seen in some AGN. Possible factors include reddening by dust on torus or galaxy scales and contamination from stellar light in the host galaxy. The reddened Type-1 AGN composite spectrum, with its flatter slope and stronger stellar features, suggests a significant contribution from both dust extinction and host galaxy light. In contrast, the blue Type-1 AGN spectrum is less affected by these factors, showing a steeper continuum and weaker stellar absorption.

\begin{figure}[t]
    \centering
    \includegraphics[scale=0.43]{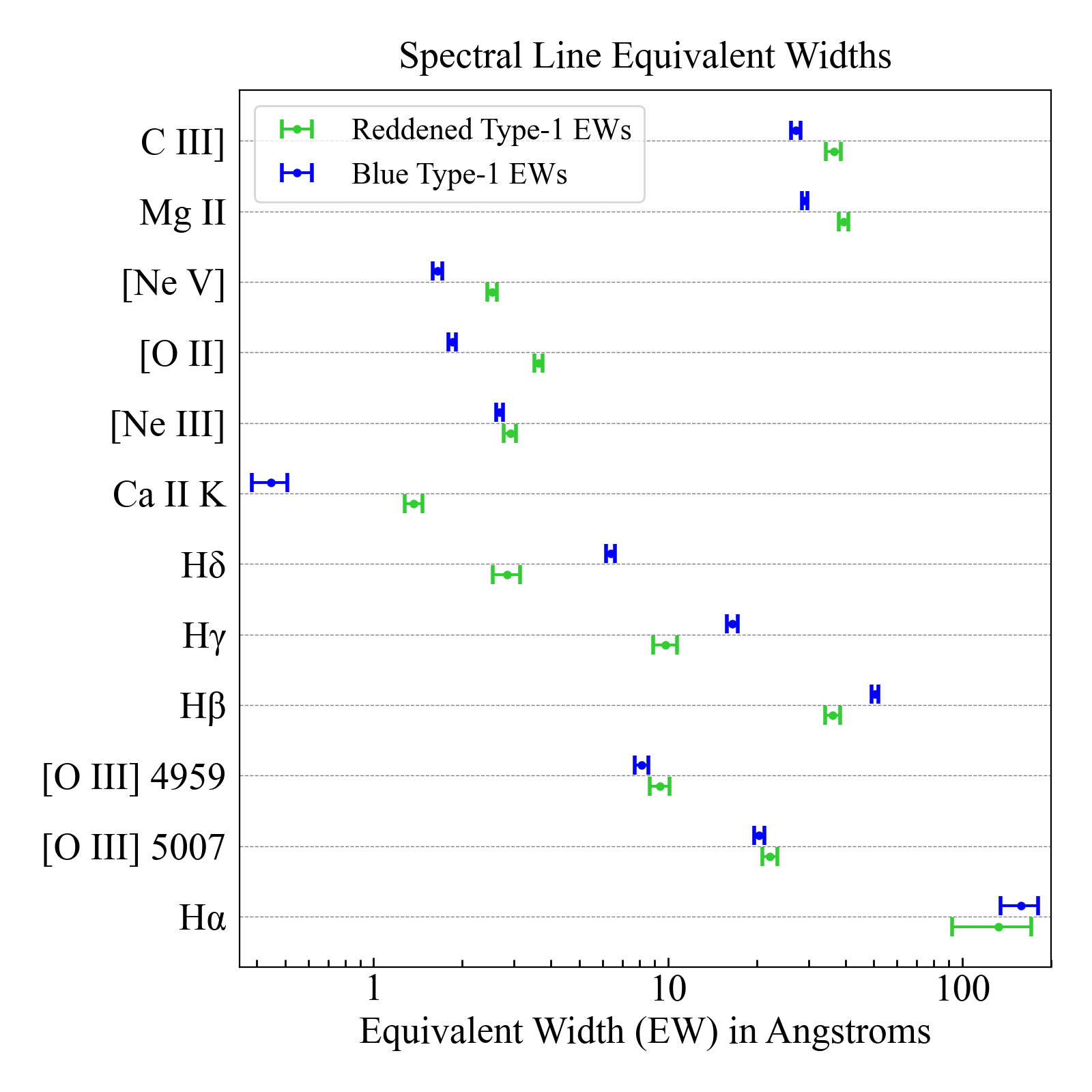}
    \caption{Equivalent widths (EWs) of prominent emission lines in blue and reddened Type-1 AGN (blue and green markers). We perform bootstrap analysis with 1000 realizations to get the 68\% confidence level interval. EWs are plotted as absolute values in log-scale for easier display.
    \label{fig:equivalent_widths}}
\end{figure}

\subsection{Spectral Line Equivalent Widths} \label{subsec:equivalent_widths}

We aim to compare the strength of emission and absorption lines in both classes. To do this, we need to make sure that the relative fluxes between the lines are preserved when stacking the spectra. Therefore, we calculate an alternate composite using the median instead of the geometric mean, following a method similar to \citealt{vandenberk2001}. We follow the same process as before, shifting the spectra to the rest frame and rebinning them to a common resolution of 1 \(\mathrm{\mathring{A}}\). However, in this case, each spectrum is scaled by its median flux density, and in the final step we take the median in each wavelength bin. With this new composite spectrum we can calculate the equivalent widths (EWs) of several prominent emission lines and estimate their uncertainty by performing a bootstrap analysis with 1000 realizations as before. The equivalent width values and their 68\% confidence level are reported in Table \ref{tab:lines}.

\renewcommand{\arraystretch}{1.5}
\begin{deluxetable}{cccc} 
\tablecaption{Spectral Line Equivalent Widths \label{tab:lines}}
\tablehead{
    \colhead{ID} & \colhead{$\mathrm{\lambda_{lab}~[\mathring{A}]}$} & \colhead{Blue $\mathrm{W_{\lambda}~[\mathring{A}]}$} & \colhead{Reddened $\mathrm{W_{\lambda}~[\mathring{A}]}$} 
}
\startdata
C III$]$     & 1908.73 & $-27.22^{-1.01}_{+1.00}$ & $-36.57^{-2.21}_{+2.15}$ \\ 
Mg II        & 2798.75 & $-29.12^{-0.61}_{+0.60}$ & $-39.53^{-1.49}_{+1.50}$ \\  
$[$Ne V$]$   & 3426.84 &  $-1.65^{-0.06}_{+0.06}$ &  $-2.53^{-0.09}_{+0.10}$ \\  
$[$O II$]$   & 3728.48 &  $-1.85^{-0.05}_{+0.05}$ &  $-3.63^{-0.12}_{+0.11}$ \\  
$[$Ne III$]$ & 3869.85 &  $-2.68^{-0.08}_{+0.07}$ &  $-2.91^{-0.14}_{+0.13}$ \\  
Ca II K      & 3934.78 &  $+0.45^{+0.06}_{-0.06}$ &  $+1.37^{+0.10}_{-0.09}$ \\  
H$_{\delta}$ & 4102.89 &  $-6.38^{-0.23}_{+0.20}$ &  $-2.84^{-0.30}_{+0.30}$ \\  
H$_{\gamma}$ & 4341.68 & $-16.52^{-0.67}_{+0.70}$ &  $-9.79^{-0.89}_{+0.92}$ \\  
H$_{\beta}$  & 4862.68 & $-50.49^{-1.31}_{+1.33}$ & $-36.22^{-2.08}_{+2.11}$ \\  
$[$O III$]$  & 4960.30 &  $-8.14^{-0.42}_{+0.45}$ &  $-9.38^{-0.70}_{+0.72}$ \\  
$[$O III$]$  & 5008.24 & $-20.41^{-0.76}_{+0.79}$ & $-22.17^{-1.29}_{+1.30}$ \\  
H$_{\alpha}$ & 6564.61 &$-158.27^{-23.22}_{+22.76}$&$-132.51^{-40.19}_{+38.53}$ \\ 
\enddata
\tablecomments{Negative values represent emission and positive values represent absorption.}
\end{deluxetable}

The EWs of prominent emission lines in blue and reddened Type-1 AGN show notable differences, as illustrated in Figure \ref{fig:equivalent_widths}. Narrow emission lines (e.g., [Ne V], [O II], [Ne III], [O III]) show stronger EWs in reddened Type-1 AGN compared to blue Type-1 AGN. In contrast, broad Balmer emission lines (e.g., H$_\mathrm{\alpha}$, H$_\mathrm{\beta}$, H$_\mathrm{\gamma}$, H$_\mathrm{\delta}$) have stronger EWs in blue Type-1 AGN compared to their reddened counterparts. An exception to this trend can be observed in Mg II and C III], but these broad lines have a prominent narrow-line component. The implications of these results will be discussed in \S \ref{subsec:ew_analysis}.

\subsection{Comparison between \texttt{CIGALE} fits to Blue and Reddened Type-1 AGN} \label{subsec:parameter_comparison}

\begin{figure*}[t]
    \centering
    \includegraphics[width=\textwidth]{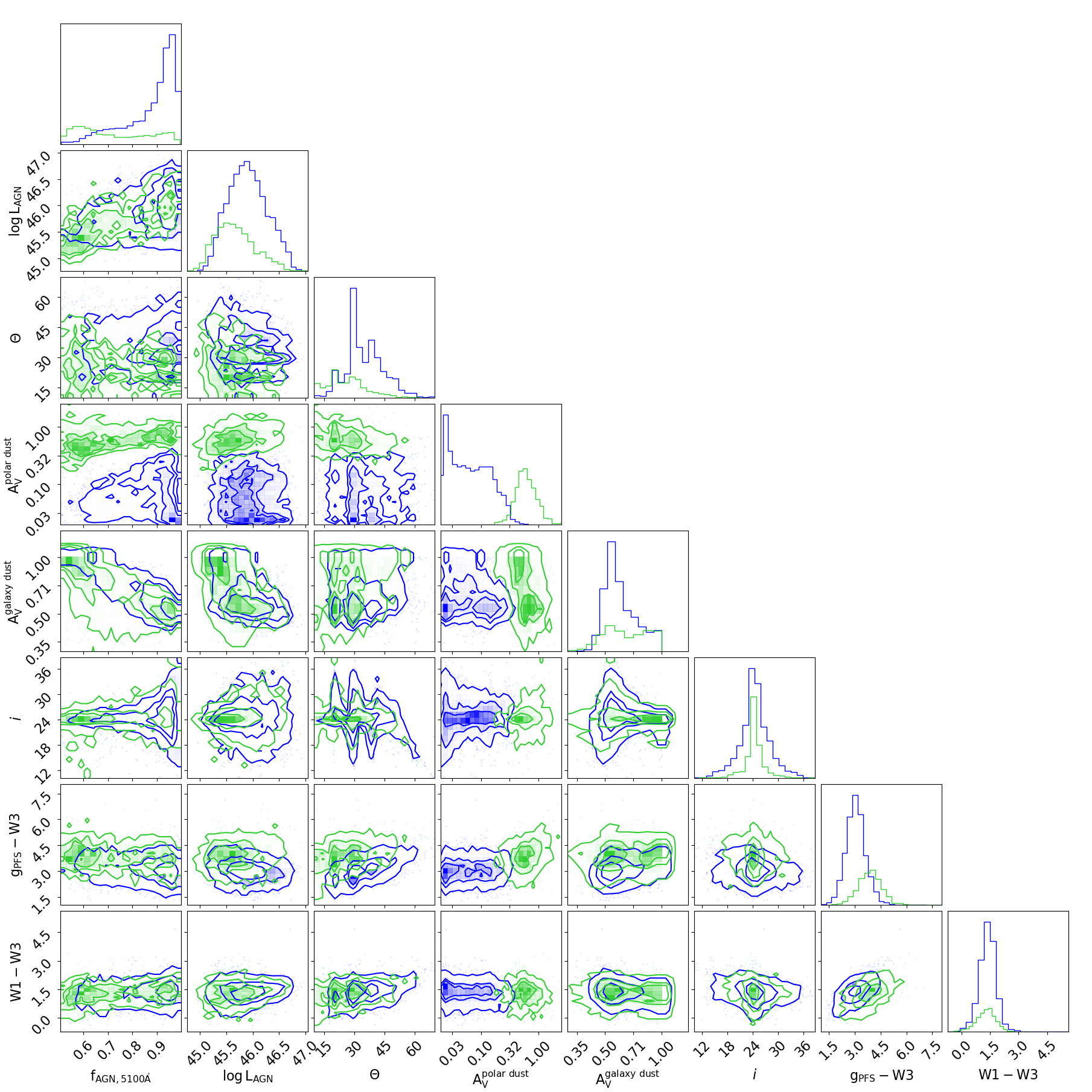}
    \caption{Corner plot showing the distribution of AGN and host galaxy parameters derived from the \texttt{CIGALE} SED fitting and photometric colors for the blue (blue) and reddened (green) Type-1 AGN subsamples. Contour levels enclose 38, 68, and 95\% of each subsample; individual objects in the low-density outskirts are shown as scatter points. The diagonal panels show the marginals of each parameter. The parameters shown are the AGN fraction at 5100~\AA\ ($\mathrm{f_{AGN,5100\mathring{A}}}$), AGN luminosity ($\mathrm{\log L_{AGN}}$), torus half-opening angle ($\Theta$), polar dust attenuation ($\mathrm{A_{V}^{polar~dust}}$), host galaxy dust attenuation ($\mathrm{A_{V}^{galaxy~dust}}$), inclination angle ($i$), and observed optical--mid-IR colors ($\mathrm{g_{PFS}-W3}$ and $\mathrm{W1-W3}$).
    \label{fig:corner}}
\end{figure*}

Using the classification described in \S \ref{subsec:alphas}, we compare the parameter distributions obtained from the SED fitting of blue and reddened Type-1 AGN. Figure \ref{fig:corner} presents a corner plot with some of the most relevant parameters fitted in the SED modeling using \texttt{CIGALE}. Table \ref{tab:medians} summarizes the median values and 68\% confidence intervals for key parameters, along with a z-test to assess the significance of their differences.

The distribution for the average equatorial optical depth of the dusty torus at $9.7\mu m$ ($\tau_{9.7}$) peaks around 7.0 for both populations, in agreement with \citet{yang2020}, suggesting similar overall dust column densities in the torus. The inclination angle distributions are also nearly identical, with median values of $i=24.3^{+3.2}_{-3.1}$ deg for blue Type-1 AGN and $i=24.3^{+1.4}_{-1.6}$ deg for reddened Type-1 AGN. The z-test result of 0.389 confirms that both distributions are statistically indistinguishable. Given that these sources are spectroscopically confirmed broad-line AGN, the fitting results align well with \citet{yang2020}, where broad lines are preferentially observed at lower inclinations.

In contrast, the torus half-opening angle shows a clear difference. While blue Type-1 AGN have a median $\Theta=33.3^{+11.1}_{-5.9}$ deg, reddened Type-1 AGN exhibit a lower median of $\Theta=25.7^{+10.1}_{-8.7}$ deg, consistent with a more compact torus structure. The z-test value of 26.574 confirms that these distributions are significantly distinct, with blue Type-1 AGN peaking around values consistent with \citet{Stalevski2016}, while reddened AGN tend toward smaller half-opening angles, meaning their tori are thinner. A similar trend is seen in polar dust extinction, where reddened Type-1 AGN show systematically higher values. The median dust extinction in $\mathrm{A_{V}}$ from polar dust is $0.60^{+0.32}_{-0.19}$ mag in reddened Type-1 AGN, compared to $0.06^{+0.10}_{-0.03}$ mag in blue Type-1 AGN, following the $\mathrm{E(B-V)}$ to $\mathrm{A_{V}}$ conversion with $\mathrm{R_{V}=3.1}$. The z-test value of -92.521 confirms the stark contrast, pointing to substantial obscuration from dust in the polar region of reddened Type-1 AGN. The implications of these results are discussed further in \S \ref{subsec:agn_structure_diff}.

The contributions from the host-galaxy dust component are broadly similar between the two populations, with both exhibiting a peak near $\mathrm{A_{V}\sim0.6~mag}$. However, reddened Type-1 AGN show a tendency toward higher obscuration ($\mathrm{A_{V}\sim0.9~mag}$), producing an apparent bimodality. If all of the extinction came solely from host-galaxy dust (i.e. large galactic scales), we would expect the AGN and stellar emission to be affected in similar ways. Instead, the need for an additional component that primarily affects AGN emission suggests that this dust is located in much smaller scales, potentially in the vicinity of the torus or broad line region, or arises from a circumnuclear dust lane. Notably, we don't find a significant correlation between the "polar dust" component and the galaxy dust component.

Reddened Type-1 AGN also exhibit redder mid-IR colors, with a median $\mathrm{g_{PFS}-W3}$ color of $3.88^{+0.64}_{-0.66}$ mag compared to $3.03^{+0.58}_{-0.52}$ mag for blue Type-1 AGN. Additionally we note that $\mathrm{g_{PFS}-W3}$ and W1-W3 colors show no clear correlation with polar dust reddening for blue Type-1 AGN, while for reddened Type-1 AGN, increasing $\mathrm{A_{V}^{polar~dust}}$ is associated with redder $\mathrm{g_{PFS}-W3}$ colors. This behavior is consistent with models where mid-IR emission originates not just from the torus but also from polar dust outflows, as proposed by \citet{honig2012, honig2013}. In this picture, radiation-driven winds lift dust into the polar region, where it absorbs and re-emits AGN radiation, contributing to both the observed reddening and enhanced mid-IR flux in reddened Type-1 AGN. The AGN contribution to the total flux at $5100$~\AA\ also differs, with blue Type-1 AGN having a median $f_{\text{AGN, 5100\AA}}$ of $0.91^{+0.05}_{-0.17}$, compared to $0.69^{+0.21}_{-0.12}$ in reddened AGN. This lower AGN fraction suggests that in reddened AGN, more of the observed optical flux comes from the host galaxy, likely due to attenuation of the AGN continuum by dust. The distributions of luminosities that come from the SED fit with \texttt{CIGALE} for blue and reddened Type-1 AGN are similar, but there is a very slight preference for reddened objects to smaller AGN luminosities by 0.25 dex. These AGN luminosities represent the sum of the observed AGN disk luminosity and dust re-emitted luminosity from the fits.

Since the half-opening angle of the torus and reddening from dust in the polar region seem to be the parameters that constrain the shape of the SED more strongly, we want to do a deeper analysis of how these two parameters are related to each other in the next section, and will discuss them in detail in \S \ref{subsec:agn_structure_diff}.

\begin{figure*}[t]
    \centering
    \includegraphics[width=\textwidth]{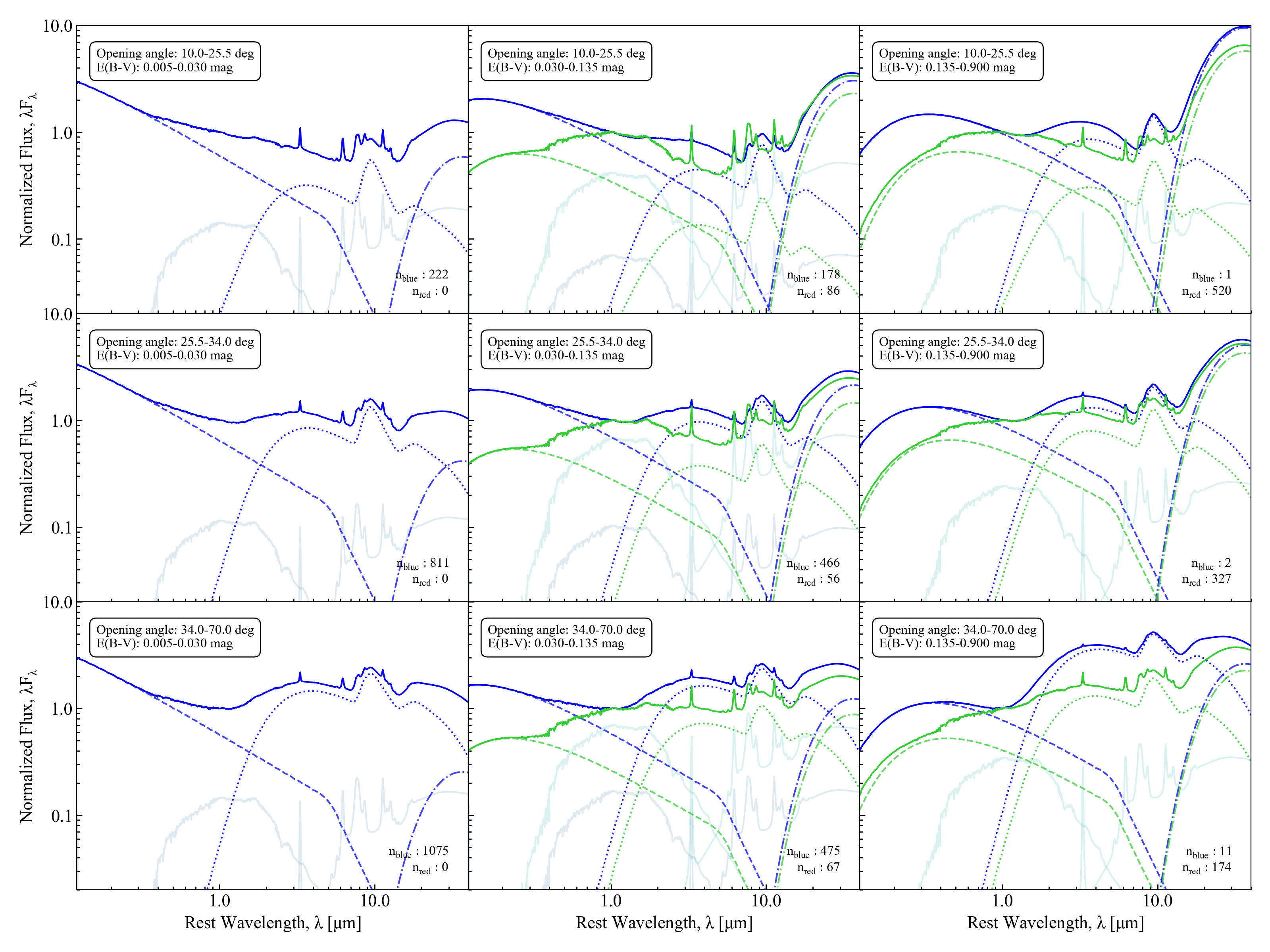}
    \caption{Effects of half-opening angle and E(B-V) polar dust reddening in the shape of the SED from optical to mid-IR.
    Half-opening angles are separated into $10.0^\circ-25.5^\circ$ (1st row), $25.5^\circ-34.0^\circ$ (2nd row), and $34.0^\circ-70.0^\circ$ (3rd row). Polar dust reddening values in E(B-V) are divided into $0.005-0.030 \mathrm{~mag}$ (1st column), $0.030-0.135 \mathrm{~mag}$ (2nd column), and $0.135-0.900 \mathrm{~mag}$ (3rd column). Blue Type-1 AGN are shown in blue and reddened Type-1 AGN are shown in green. A geometric mean of all the SED fits that fall within the values in each square are shown in solid lines. The different components of the AGN SED are also plotted. A dashed line represents the reddened emission from the accretion disk, a dotted line represents the emission from the dusty torus, and a dot-dashed line represents the emission from polar dust. Solid lines in light-blue and light-green represent the emission from the host galaxy. The number of sources in each square is shown in the bottom right corner.
    \label{fig:grid}}
\end{figure*}

\subsection{Relation between Polar Dust Reddening and Half-Opening Angle of the Torus} \label{subsec:ebv_vs_oa}

To explore the relationship between the dusty torus half-opening angle, that relates to the torus height, and extinction due to dust in the polar direction, we construct a grid that samples low, intermediate, and high values of both parameters. This approach allows us to examine variations in the SED shape and evaluate the sensitivity of these parameters to the available photometry. As shown in Figure \ref{fig:grid}, we group sources into three bins of torus half-opening angles ($10.0^\circ - 25.5^\circ$, $25.5^\circ - 34.0^\circ$, and $34.0^\circ - 70.0^\circ$) and three bins of polar dust reddening ($E(B - V)=0.005 - 0.030$, $0.030 - 0.135$, and $0.135 - 0.900$ mag). For each combination, we compute the geometric mean of the best-fit SEDs, with blue and reddened Type-1 AGN plotted in blue and green, respectively. The total SED is plotted as a solid line, while the individual AGN components are shown as a dashed line for the accretion disk, a dotted line for the dusty torus, and a dash-dotted line for polar dust. Contributions from the host galaxy are shown as solid light-blue and light-green lines for blue and reddened Type-1 AGN, respectively.

The most suitable bins for comparing blue and reddened Type-1 AGN are those with intermediate polar-dust reddening ($\mathrm{E(B-V)}=0.030-0.135 \mathrm{~mag}$ or $\mathrm{A_{V}=0.093-0.418~mag}$), where both populations are well represented. In contrast, bins with the highest polar-dust reddening contain very few blue Type-1 AGN, especially at small half-opening angles, and no statistical comparisons can be done between the two populations in a given bin. In those cases, the key point is simply that the number of blue versus reddened Type-1 AGN in each bin differs dramatically.

As highlighted in Figures \ref{fig:corner} and \ref{fig:grid} above, reddened Type-1 AGN are preferentially located at small torus half-opening angles ($10.0^\circ - 25.5^\circ$) and higher polar-dust reddening ($\mathrm{E(B-V)>0.135~mag}$ or $\mathrm{A_{V}>0.418~mag}$), whereas blue Type-1 AGN are more common at larger half-opening angles ($34.0^\circ - 70.0^\circ$) and low reddening ($\mathrm{E(B-V)<0.030~mag}$ or $\mathrm{A_{V}<0.093~mag}$). This separation is clearest in the highest reddening bin, which is dominated by reddened sources.

The wavelength range from $2.0-10.0~\mathrm{\mu m}$ is dominated by the mid-IR emission from the dusty torus in blue Type-1 AGN. A similar trend is seen in reddened Type-1 AGN, but in this case the contribution from host-galaxy dust decreases as the effects of polar-dust extinction increase. At longer wavelengths $(>10.0~\mathrm{\mu m})$, the situation becomes more complex: reddened Type-1 AGN exhibit weaker emission from a thinner dusty torus, but stronger emission from both the host-galaxy dust and the polar-dust component. These competing effects lead to overall comparable emission levels in this wavelength range for both samples.  In the UV–optical region, reddened Type-1 AGN exhibit the expected steeper slopes caused by stronger extinction. The ratio of torus to polar dust emission in the near-IR varies with half-opening angle: at larger angles, torus emission dominates, while at smaller angles, polar dust contributes more significantly. Similarly, as polar dust reddening increases, the SEDs exhibit a redder UV-optical slope and stronger near-IR excess, reinforcing the link between low torus half-opening angles (small torus height) and high polar dust extinction.

The host galaxy contribution is consistently stronger in the SEDs of reddened Type-1 AGN compared to blue Type-1 AGN within a given bin. This is consistent with greater attenuation of the nuclear emission in these systems, allowing the host galaxy light to contribute more significantly to the observed flux. The torus component is weaker in reddened Type-1 AGN SEDs compared to blue Type-1 AGN, particularly in bins with low half-opening angles, supporting the interpretation that these objects preferentially occupy parameter spaces with smaller torus half-opening angles.

\begin{deluxetable}{cccc} 
\tablecaption{Distribution of Relevant Parameters \label{tab:medians}}
\tablehead{
    \colhead{Parameter} & \colhead{Blue Type-1} & \colhead{Reddened Type-1} & \colhead{Z-Test}
}
\startdata
$i$ & $24.3^{+3.2}_{-3.1}$ deg & $24.3^{+1.4}_{-1.6}$ deg & 0.389 \\
$\Theta$ & $33.3^{+11.1}_{-5.9}$ deg & $25.7^{+10.1}_{-8.7}$ deg & 26.574 \\
$\mathrm{A_{V}^{polar~dust}}$ & $0.06^{+0.10}_{-0.03}$ mag & $0.60^{+0.32}_{-0.19}$ mag & -92.521 \\
$\mathrm{A_{V}^{galaxy~dust}}$ & $0.57^{+1.76}_{-0.99}$ mag & $0.62^{+2.09}_{-0.86}$ mag & -7.264 \\
$\mathrm{g_{PFS} - W3}$ & $3.03^{+0.58}_{-0.52}$ mag & $3.88^{+0.64}_{-0.66}$ mag & -39.623 \\
$f_{\text{AGN, 5100\AA}}$ & $0.91^{+0.05}_{-0.17}$ & $0.69^{+0.21}_{-0.12}$ & 37.886 \\
\enddata
\tablecomments{Values are the median and 68\% confidence intervals for key parameters that result from the SED fitting with \texttt{CIGALE}, along with a z-test to assess the significance of their differences.}
\end{deluxetable}

\section{Discussion} \label{sec:discussion}

We present a detailed analysis of the spectral energy distributions (SEDs) of Type-1 quasars to investigate the physical origin of their reddening. By modeling the SEDs from the near-UV to the mid-IR, we have been able to constrain key properties of the sources of reddening and their radial extent. Here, we discuss evidence for structural differences between the two classes on sub-pc scales.

\subsection{Intrinsic differences in AGN Structure} \label{subsec:agn_structure_diff}

In the context of AGN structure, several studies have suggested that most reddened Type-1 AGN are found in merging galaxies. For instance, some studies suggest that mergers between gas-rich galaxies can trigger both star formation and AGN activity, leading to significant dust and gas obscuration within the host galaxy \citep{sanders1988,hopkins2006}. This scenario implies that the additional dust causing the reddening in these AGN acts as a screen from the galaxy itself, rather than being intrinsic to the AGN's central structure. Observational evidence from \citealt{urrutia2008} and \citealt{glikman2015} shows higher fractions of merger characteristics in reddened Type-1 AGN, supporting this argument. If the primary difference between blue and reddened Type-1 AGN was merely a dust screen from the host galaxy, we would not expect to find significant differences in the intrinsic structure of the dusty torus surrounding the AGN, such as in the half-opening angle across both populations.

However, our \texttt{CIGALE} fits reveal clear differences in the torus structures of blue and reddened Type-1 AGN. Reddened Type-1 AGN tend to have smaller torus half-opening angles, suggesting a more compact torus. For objects with moderate reddening, where broad lines are still visible but the continuum is clearly reddened, it is unlikely that the torus is the main cause of the obscuration in the rest-frame optical/UV. In these cases, \texttt{CIGALE} adds an extra reddening component called “polar dust,” which acts as a dust screen along the line of sight to the accretion disk. This dust could, in principle, be located anywhere along the line of sight, from the nucleus to the host galaxy. In the composite SEDs shown in Figure \ref{fig:grid}, the dusty torus component is plotted as blue and green dotted lines for blue and reddened Type-1 AGN, respectively. The torus emission in the $\mathrm{2-30 , \mu m}$ range is weaker in reddened Type-1 AGN compared to the blue ones. This is also seen in the parameter distributions from our fits (Figure \ref{fig:corner}), which show that reddened Type-1 AGN favor smaller half-opening angles, resulting in a thinner torus with less mid-IR emission. Although the polar dust component is flexible in location, we will discuss below why we think the dust causing the reddening is likely nuclear, close to the AGN.

Reddening from dust in the polar direction seems to be slightly correlated with the half-opening angle of the dusty torus in reddened Type-1 AGN. There seems to be a link between the amount of extinction and a need for a larger region to fill with relatively diffuse dust. The peak around $30^\circ$ for half-opening angles in blue Type-1 AGN permits an optimal balance: it provides a clear line of sight for observing broad emission while minimizing the area available for polar dust extinction. Reddened Type-1 AGN, on the other hand, seem to favor lower half-opening angles where the dusty torus is thin, leading to substantial dust reddening in the wide region covered by polar dust. At higher half-opening angles, the amount of polar dust is restricted by geometry, resulting in reduced polar dust reddening.

\subsection{Narrow-line Equivalent Widths argue for torus-scale dust} \label{subsec:ew_analysis}

The differences in the equivalent widths (EWs) of emission lines between blue and reddened Type-1 AGN provide additional evidence for distinct dust structures on sub-parsec scales. Specifically, we observe a notable trend: the EWs of narrow emission lines are systematically higher in reddened Type-1 AGN, whereas the EWs of broad emission lines, particularly the Balmer series, tend to be lower compared to their blue counterparts (see Table \ref{tab:lines}). Broad emission lines, such as H$\alpha$, H$\beta$, Mg II, and C III], are expected to originate in the broad-line region (BLR), a compact zone near the supermassive black hole where gas is exposed to the intense ionizing radiation from the accretion disk. The BLR gas is generally believed to be in hydrostatic equilibrium, meaning it is gravitationally bound and not participating in an outflow. This distinguishes it from dust in the polar direction, which interferometric observations suggest is often associated with outflowing material \citep{asmus2016,stalevski2017}. In contrast, narrow emission lines, including [O III], [Ne III], [O II], and [Ne V], arise from the narrow-line region (NLR), a more extended region farther from the central source. We note that Mg II and C III] deviate somewhat from the broad Balmer line behavior, likely because these lines have a stronger narrow-line component that contributes to their total flux.

Because the EW is defined as the ratio of line flux to continuum flux, dust along the line of sight can strongly affect its value. Dust obscures the bright continuum from the accretion disk and the broad-line region (BLR) more than the narrow-line region (NLR), which is farther away from the central black hole and more extended. As a result, the increase in narrow-line EWs in reddened Type-1 AGN suggests that the dust is not spread uniformly across the entire galaxy. If it were, it would dim both the continuum and the NLR equally, and the narrow-line EWs would stay the same. Instead, the stronger narrow-line EWs point to dust concentrated closer to the AGN, likely on scales $\mathrm{\sim0.1-10~pc}$, which dims the continuum and broad-lines, while leaving the more distant NLR mostly unaffected. 

This interpretation aligns with the SED-based results presented in \S \ref{subsec:ebv_vs_oa}, which indicate that the reddening in these objects is associated with low torus half-opening angles (small torus heights) and increased polar dust extinction. Taken together, these findings suggest that the dust responsible for the attenuation resides on parsec scales. This structure obscures the nuclear continuum along certain lines of sight while permitting a relatively unobstructed view of the NLR.

Likewise, the preference for low half-opening angles in reddened Type-1 AGN seems to reflect real physical differences between the two classes, as suggested by the trends in our data. While the half-opening angle is a key factor, another parameter could be driving this connection. One possibility is the accretion rate onto the supermassive black hole, $\dot{M}$. AGN with higher $\dot{M}$ are thought to produce stronger radiation-driven outflows, which could push down on the torus, making it more compact. At the same time, these outflows could lift dust into the polar region, increasing the amount of material along the line of sight and leading to stronger reddening \citep{honig2013,calistro2021}. This idea is similar to the dusty wind model proposed by \citet{honig2012, honig2013}, which suggests that much of the mid-IR emission in AGN comes from dust in outflows rather than just the torus. If these outflows get stronger at higher accretion rates, they could explain both the increased polar dust reddening and the smaller torus half-opening angles we see in reddened Type-1 AGN. This points to a connection between accretion rate, torus structure, and polar dust, which could be tested with high-resolution IR observations that map out dust in both the torus and the polar region.

It is worth noting that if the polar dust responsible for the reddening resides on parsec scales, it should obscure the continuum emission from the accretion disk and the BLR in a similar way. In that case, we would expect comparable broad Balmer line EWs in both blue and reddened Type-1 AGN. However, as shown in Table \ref{tab:lines} and Figure \ref{fig:equivalent_widths}, the broad Balmer line EWs are systematically smaller in reddened Type-1 AGN. We address this peculiarity in the next section, where we discuss how larger host-galaxy contributions to the continuum in reddened Type-1 AGN naturally suppress their broad Balmer line equivalent widths.

\subsection{Differences in galaxy contribution} \label{subsec:galaxy_contributions}

Stronger host galaxy contributions are evident in the composite spectrum of reddened Type-1 AGN (Figure \ref{fig:composite}). In particular, the Ca II H and K absorption lines, along with the higher-order Balmer series absorption features, are noticeably more pronounced compared to those in blue Type-1 AGN. These absorption lines are characteristic of intermediate-age stellar populations and indicate that starlight from the host galaxy contributes more significantly to the integrated spectrum in reddened Type-1 AGN. This is consistent with the interpretation that the AGN continuum in reddened sources is more heavily attenuated by dust, allowing the underlying stellar absorption features to emerge more clearly. In contrast, in blue Type-1 AGN, the bright, unobscured AGN continuum largely outshines the host galaxy’s stellar light, diluting these absorption features. We also note that Fe II emission, which is prominent in the optical spectra of Type-1 AGN, appears to be stronger in blue Type-1 AGN compared to their reddened counterparts. This Fe II emission can blend with the higher-order Balmer lines, filling in and diluting the absorption features associated with stellar populations in the host galaxy.

The results from the SED fitting analysis using \texttt{CIGALE} (Figure \ref{fig:grid}) further reinforce this picture. As discussed in \S \ref{subsec:ebv_vs_oa}, the fitted host galaxy contribution is stronger in reddened Type-1 AGN across the entire parameter space. Together, the spectral and SED-based evidence highlights that host galaxy starlight contributes more visibly to the observed SEDs of reddened Type-1 AGN compared to their blue counterparts.  This increased visibility of the galaxy can be interpreted as a consequence of the attenuation of the nuclear emission by dust, particularly polar dust, which reduces the AGN’s dominance over the host light.

We interpret the larger host-galaxy contribution to the observed continuum in reddened Type-1 AGN, revealed by the stronger Ca II H and K absorption features, as a potential explanation for the systematic decrease in broad Balmer-line equivalent widths toward the higher shell transitions. If the host galaxy adds extra continuum light primarily in reddened Type-1 AGN, this would dilute the broad Balmer line EWs, as well as adding additional Balmer absorption. Confirming this scenario requires higher-quality data with simultaneous fitting of deep optical–NIR spectra and photometry to more cleanly separate the AGN continuum, the host-galaxy light, and the effects of dust on each component. The Prime Focus Spectrograph (PFS; \citealt{sugai2015,tamura2024}) on the Subaru Telescope is well suited for this task. Its wide wavelength coverage ($0.38-1.26~\mu m$), high sensitivity, and large multiplexing will allow deep, uniform spectra for large AGN samples. With PFS-quality data, we can trace the full Balmer series more precisely, measure host-galaxy absorption features more reliably, and directly test whether the stronger host contribution drives the EW trends seen in reddened Type-1 AGN. These observations would place tighter constraints on the central engine structure and clarify the physical origin and distribution of dust in these sources.

\section{Conclusions}\label{sec:conclusions}

Based on our CIGALE fits to blue and reddened broad-line AGN, we find multiple lines of evidence that the dust extinction in reddened Type-1 AGN originates on parsec or sub-parsec scales. Narrow emission lines have systematically stronger equivalent widths in reddened AGN compared to their blue counterparts, suggesting that while the continuum and broad-line region are affected by obscuration, the narrow-line region remains largely unobscured. Additionally, our SED modeling reveals structural differences between the two populations, with reddened AGN exhibiting smaller torus half-opening angles (thinner torii), with a median of $25.7^{+10.1}_{-8.7}$~deg, compared to $33.3^{+11.1}_{-5.9}$~deg for blue Type-1 AGN. This may indicate a connection between polar dust and torus geometry, where the reddening is associated with a polar outflow that compresses the torus, leading to a smaller half-opening angle. We also find that reddened AGN have systematically higher polar dust extinction, with a median $\mathrm{A_V=0.60^{+0.32}_{-0.19}~mag}$, compared to $\mathrm{A_V=0.06^{+0.10}_{-0.03}~mag}$ for blue Type-1 AGN. The increased host galaxy contribution in reddened AGN also appears to be a secondary effect, resulting from the attenuation of the nuclear emission rather than an intrinsic difference in host properties.

Future observations with high spatial and spectral resolution, particularly with JWST, VLT and PFS, will be crucial for mapping the distribution of this dust and confirming its origin. High-angular-resolution imaging will help distinguish between nuclear and galaxy-scale obscuration, while spectroscopic studies will provide better constraints on the dust properties and the contributions from host galaxy. At the same time, expanding to larger AGN samples that span a broader color space, potentially by relaxing the SDSS spectroscopic requirement and using photometric data from HSC instead of synthetic photometry from SDSS spectra, will allow for a more complete census of AGN populations. This combination of targeted high-resolution studies and large statistical samples will provide a clearer picture of how polar dust, torus structure, and nuclear obscuration evolve with black hole growth and quasar evolution.

\section{Acknowledgments}

We thank the anonymous referee for their constructive and insightful comments that have significantly improved the clarity of our work.

This material is based upon work supported by the National Science Foundation (NSF) under Cooperative Agreement Nos. 1647375 and 1647378, including the National Astronomy Consortium (NAC). The NAC is made possible through the generous support of the National Radio Astronomy Observatory (NRAO) and the NSF. The NRAO is a facility of the NSF operated under cooperative agreement by Associated Universities, Inc. 

This publication makes use of data products from the Wide-field Infrared Survey Explorer, which is a joint project of the University of California, Los Angeles, and the Jet Propulsion Laboratory/California Institute of Technology, funded by the National Aeronautics and Space Administration.

This paper makes use of data collected at the Subaru Telescope, which is operated by the Subaru Telescope and Astronomy Data Center at the National Astronomical Observatory of Japan.

The Hyper Suprime-Cam (HSC) collaboration includes the astronomical communities of Japan and Taiwan, and Princeton University. The HSC instrumentation and software were developed by the National Astronomical Observatory of Japan (NAOJ), the Kavli Institute for the Physics and Mathematics of the Universe (Kavli IPMU), the University of Tokyo, the High Energy Accelerator Research Organization (KEK), the Academia Sinica Institute for Astronomy and Astrophysics in Taiwan (ASIAA), and Princeton University.

Funding for the Sloan Digital Sky Survey IV has been provided by the Alfred P. Sloan Foundation, the U.S. Department of Energy Office of Science, and the Participating Institutions. SDSS-IV acknowledges support and resources from the Center for High Performance Computing  at the University of Utah. The SDSS website is www.sdss4.org.

SDSS-IV is managed by the Astrophysical Research Consortium for the Participating Institutions of the SDSS Collaboration including the Brazilian Participation Group, the Carnegie Institution for Science, Carnegie Mellon University, Center for Astrophysics | Harvard \& Smithsonian, the Chilean Participation Group, the French Participation Group, Instituto de Astrof\'isica de Canarias, The Johns Hopkins University, Kavli Institute for the Physics and Mathematics of the Universe (IPMU) / University of Tokyo, the Korean Participation Group, Lawrence Berkeley National Laboratory, Leibniz Institut f\"ur Astrophysik Potsdam (AIP),  Max-Planck-Institut f\"ur Astronomie (MPIA Heidelberg), Max-Planck-Institut f\"ur Astrophysik (MPA Garching), Max-Planck-Institut f\"ur Extraterrestrische Physik (MPE), National Astronomical Observatories of China, New Mexico State University, New York University, University of Notre Dame, Observat\'ario Nacional / MCTI, The Ohio State University, Pennsylvania State University, Shanghai Astronomical Observatory, United Kingdom Participation Group, Universidad Nacional Aut\'onoma de M\'exico, University of Arizona, University of Colorado Boulder, University of Oxford, University of Portsmouth, University of Utah, University of Virginia, University of Washington, University of Wisconsin, Vanderbilt University, and Yale University.

\facilities{WISE, HSC, SDSS}

\software{\texttt{CIGALE} \citep{burgarella2005,noll2009,boquien2019,yang2020,yang2022}, \texttt{Astropy} \citep{astropy:2013, astropy:2018, astropy:2022}, \texttt{SpectRes} \citep{carnall2017}}, \texttt{Specutils} \citep{earl2023}, \texttt{Numpy} \citep{harris2020}, \texttt{SciPy} \citep{scipy2020}.

\bibliography{sample631}{}
\bibliographystyle{aasjournal}
\end{document}